\pdfoutput=1
\documentclass{IEEEcsmag}
\usepackage[utf8]{inputenc}
\usepackage[dvipsnames]{xcolor}
\usepackage{amsmath}
\usepackage{mathtools, cuted}
\usepackage{bm}
\usepackage{hyperref}
\usepackage{comment}
\usepackage{graphics} 
\usepackage{graphicx}
\usepackage[frozencache=true,cachedir=mintedcache]{minted}
\usepackage[justification=centering]{caption}

\hypersetup{
    colorlinks=true,
    linkcolor=blue,
    filecolor=magenta,      
    urlcolor=blue,
    pdftitle={insert title},
    pdfpagemode=FullScreen,
    }

\renewcommand{\vec}[1]{\bm{#1}}
\newcommand{\red}[1]{\textcolor{red}{#1}}
\newcommand{\blue}[2]{\textcolor{blue}{#1}}

\begin{document}

%\sptitle{Department: Head}
%\editor{Editor: Name, xxxx@email}

\title{OpenACC Acceleration of an Agent-Based Biological Simulation Framework}

\author{Matt Stack}
\affil{Department of Computer and Information Sciences, University of Delaware. Newark, DE USA\\
NVIDIA Corporation}

\author{Paul Macklin}
\affil{Department of Intelligent Systems Engineering, Indiana University. Bloomington, IN USA}

\author{Robert Searles}
\affil{NVIDIA Corporation}

\author{Sunita Chandrasekaran}%\thanks{Corresponding author: schandra@udel.edu}}
\affil{Department of Computer and Information Sciences, University of Delaware. Newark, DE USA}

%\markboth{Department Head}{Paper title}
\maketitle

%\chapterinitial{Introduction and Motivation}
\section{Introduction and Motivation}
\begin{figure*}[h]
    \centering
    \includegraphics[width=0.99\textwidth]{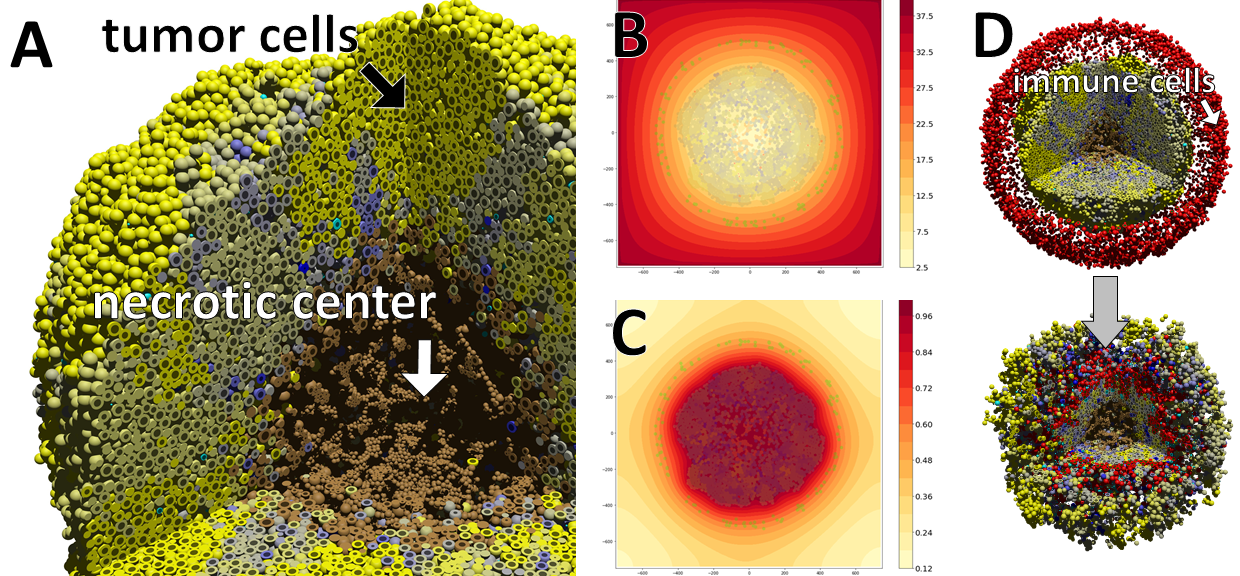}
    \caption{\textbf{Typical PhysiCell model.} The ``cancer-immune-sample'' is a typical example of a complex  biological system modeled with the PhysiCell framework. Tumor cells (\textbf{A}: colored from blue to yellow by aggressiveness) grow, divide, and die based upon local availability of oxygen that diffuses from the computational boundary. 
    Tumor cell consumption of oxygen leads to formation of oxygen gradients (\textbf{B}) and eventual necrosis (death) in the center (brown cells). Tumor cells also release a diffusible immunostimulatory factor (\textbf{C}) that attracts immune cells (red), who move by biased random migration towards tumor cells, attach, and preferentially kill highly immunogenic tumor cells (yellow) (\textbf{D}). Readers can interactively run a 2D version of this model in     a web browser at \url{https://nanohub.org/tools/pc4cancerimmune}.}
    \label{fig:physicell_example}
\end{figure*}

Computational biology has increasingly turned to agent-based modeling---which represents individual biological cells as discrete software agents---to explore complex biological systems where many cells interact through exchange of mechanical forces, exchange of diffusing chemical factors, and other biomechanical feedback \cite{JCO-review}. Biological diffusion (diffusion, decay, secretion, and uptake) is a key driver of biological tissues. Blood vessels release nutrients (oxygen, 
glucose, and other key metabolites) that diffuse through tissues to be consumed by 
cells and then absorb diffusible waste products, while cells
secrete and absorb diffusible chemical factors to communicate and coordinate 
their behaviors \cite{macklin10_cup2,macklin16_springer_TME_sysbio}. See Figure \ref{fig:physicell_example} for a typical 3-D simulation model.

Therefore, most modern agent-based biological modeling systems are \emph{hybrid}: they combine discrete cell agents with partial differential equations (PDEs) to represent biological transport, such as 

\begin{equation}
\begin{aligned}
\frac{\partial \bm{\rho}}{\partial t} = 
\bm{D} \nabla^2 \bm{\rho} - 
\bm{\lambda}\bm{\rho} +\\ 
\sum_{\textrm{cells }i} 
\delta\left( \vec{x}-\vec{x}_i \right) V_i 
\biggl(
\bm{S}_i (\bm{\rho}_i^* - \bm{\rho} )  - \bm{U}_i \bm{\rho} 
 \biggr)
 \label{PDE}
 \end{aligned}
\end{equation}

where $\bm{\rho}$ is a vector of diffusible substrates, and 
each cell agent $i$ has position $\bm{x}_i$ and volume $V_i$, a vector of 
secretion rates $\bm{S}_i$ and uptake rates $\bm{U}_i$, and a ``target'' 
extracellular substrate vector $\bm{\rho}^*i$ \cite{ghaffarizadeh15_bioinformatics}. (Vector-vector products are taken 
element-wise, and $\delta$ is the Dirac delta function.) Numerical stability and accuracy require that these PDEs be solved with relatively small step sizes $\Delta t$, making solution of biological diffusion PDEs a rate-limiting step in hybrid 
agent-based biological models that can limit the maximum size and duration of simulations. This, in turn, can hinder high-throughput simulation model exploration (e.g., 
\cite{EMEWS1,EMEWS2}) or newer model calibration techniques like approximate Bayesian computation \cite{toni_approximate_2009,da_costa_model_2018}, both of which require running thousands or millions of 
simulations quickly. 

GPU computing can vastly accelerate the diffusion and decay operators 
in the Equation \ref{PDE}, but the tight coupling with a cell-based sources--which dynamically move and change their sizes and rate constants--makes overall solution more challenging. Moreover, the number of discrete cell agents can change by orders of magnitude over the course of a cancer simulation. Recent examples (e.g., \cite{yalla}) have confirmed the potential for GPU-accelerated agent-based models, but to date have generally been limited to simplified systems that do not include complex intracellular-level dynamics (e.g., by coupling Boolean network models \cite{PhysiBoSS}) or complex cell-cell interactions and other rules. 

In this paper, we carefully utilize OpenACC, a directive-based programming model to accelerate the diffusion portion in Equation \ref{PDE} in \emph{PhysiCell}  \cite{physicell}, a cross-platform agent-based biosimulation framework that has 
been adopted in cancer \cite{liver_mets,breast_hypoxia}, infectious diseases \cite{maraviroc_covi19,covid19_coalition} and other 
complex biological problems. Prior to this work, PhysiCell had been optimized for multicore simulations with OpenMP. With default settings, PhysiCell advances the numerical solution of Equation \ref{PDE} with 10 diffusion 
steps ($\Delta t_\textrm{diff}$) before advancing the cell-cell mechanical 
interactions by one mechanics step ($\Delta t_\textrm{mech}$); cell biological 
processes (e.g., cell cycle progression, death, and ``decision making'' are advances on a much slower cell time scale $\Delta t_\textrm{cell} \sim 60 \Delta t_\textrm{mech}$. This separation of time scales allowed us to prioritize GPU optimization to the biological diffusion solver. 

Using NVIDIA HPC SDK OpenACC 21.3, we demonstrate an almost 40x speedup using managed memory on the state-of-the-art NVIDIA Ampere
100 (A100) GPU compared to a serial AMD EPYC core 7742 for a 360 simulated minutes input dataset.\sloppy %adding sloppy format to fit in the boundaries
We also demonstrate 9x speedup on the 64 core AMD EPYC 7742  multicore platform using NVIDIA HPC SDK OpenACC 21.3. 
By using OpenACC for both the CPUs and the GPUs, we maintain a single source code base, thus creating a portable yet performant solution. 

This is a critical step towards a portable GPU acceleration of a detailed agent-based simulation platform for complex biological systems. With the simulator's most significant computational bottleneck significantly reduced, we can look towards continuing cancer simulations over much longer times. 
It also represents a key step towards converting this code from purely CPU-based towards an MPI+X paradigm. In doing so, large 3-D simulation domains are decomposed into smaller subdomains residing on individual HPC nodes 
with MPI data exchange. In each compute node's subdomain, 
cell ``logic'' continues to be parallelized for multicore SMP with OpenMP and critical \emph{low-level} mechanics like biological diffusion and cell-cell mechanical interactions are parallelized on the GPU (or other specialized accelerators) via OpenACC. This will help grow the code to allow 3-D multiscale simulations of cancer and diseases in unprecedented detail on emerging leadership class high performance computing resources: a critical step towards advancing 
cancer patient digital twins that can simulate clinically significant tumors fast enough to be clinically actionable for patient care decisions \cite{precision_oncology}.

% Sunita and Paul todo
%here we need to focus on the intro and motivation both from the science and the computational view

%\section{Related Work}
%\input{related}
% Sunita and Paul todo
%here we need one paragraph on openacc and the codes it has been targeting - I can take this todo
%we need related work ideas from Paul on physicell - any OpenMP work etc., there has been no GPU work

\section{Design and Implementation}
This section narrates the design and implementation choices we employed while parallelizing and accelerating PhysiCell on a heterogeneous system. 

\subsection{Directive-Based Programming with OpenACC}
OpenACC is a performance-portable, directive-based parallel
programming model that targets modern heterogeneous HPC
hardware~\cite{openaccspec}. 
Several real-world applications and top HPC applications use OpenACC and for the same reasons we felt confident to use OpenACC for PhysiCell. These OpenACC applications include ANSYS~\cite{ansys}, Gaussian~\cite{quantum}, COSMO~\cite{fuhrer2014towards}, VASP~\cite{maintz2018strategies}, MPAS-A~\cite{yang2019accelerating} and Icosahedral non-hydrostatic (ICON)~\cite{sawyer2014towards}, MPAS WSM6~\cite{kim2021gpu} and ACME~\cite{norman2015experiences}. 

%Having said that, 

The compiler directives that it provides can be used to annotate loops and regions of code that exhibit parallelism. The programmer can also specify a target at compile time to let the compiler know what type of hardware is being targeted by the annotated code regions. Supported hardware includes multicore CPUs (both x86 and ARM), as well as various types of accelerators, including NVIDIA GPUs, AMD GPUs, and FPGAs.

OpenACC employs a host/device execution model. This means that most of a programmer's application executes within a host thread and compute intensive parallel regions (denoted by directive annotations) are offloaded either to multiple host cores (multicore CPU) or an accelerator (such as a GPU). The OpenACC specification refers to the host thread as the \textit{host} and the parallel hardware, whether it be a multicore CPU or an accelerator, as the \textit{device}. The job of the device is to execute portions of the code that are annotated with directives.  

Since many parallel regions contain multi-dimensional loop nests and modern accelerators support multiple levels of parallelism, OpenACC provides clauses the programmer can use within a directive to describe hierarchical parallelism within their code. Accelerators and multicore CPUs both support coarse-grained parallelism: fully parallel execution across execution units with limited synchronization support. OpenACC exposes coarse-grained parallelism via the \textit{gang} clause. Most accelerators and some multicore CPUs also support fine-grained parallelism: multiple execution threads within a single execution unit. OpenACC exposes fine-grained parallelism via the \textit{worker} clause. Last, most accelerators and multicore CPUs support SIMD operations within execution units. OpenACC exposes these operations via the \textit{vector} clause. It is the job of the programmer to understand the difference between these levels of parallelism and know which portions of their code can execute in what manner in order to properly annotate their source code for parallel execution.

One defining characteristic of a host/device execution model is that device memory may be separate from host memory. For example, GPUs have their own onboard memory, which is separate from the rest of the system's memory. In such a system, the host thread is unable to read or write device memory, since it is not part of the host thread's virtual memory space, and the device is similarly unable to read or write host memory in most cases. Instead, all data movement between host and device memory must be performed by the host thread through explicit system calls. Fortunately, OpenACC provides directives for managing data movement between the host and device, and the compiler takes care of generating the appropriate system calls based on the specified target accelerator. When running on NVIDIA GPUs, OpenACC applications compiled with NVIDIA's compiler can take advantage of a feature called \textit{managed memory}, where data movement between the host and the device is handled automatically by the underlying CUDA runtime, alleviating the programmer of this responsibility. However, abdicating control of data movement to the CUDA runtime may come at a performance cost. Typically, data transfer overhead is where most parallel applications that utilize an accelerator suffer performance losses. If data movement is not efficiently managed, these losses can even be so big as to outweigh the computational benefit of using the accelerator, resulting in a slowdown of overall application performance. 

The natural question is could we have used OpenMP offloading~\cite{openmp} for this code. The offloading model is beginning to mature as we speak based on the validation and verification findings~\cite{diaz2018openmp,diaz2019analysis}. We also note that the model is being used on mini-applications~\cite{openmpdavis} and other applications including Pseudo-Spectral Direct Numerical Simulation-Combined Compact Difference (PSDNS-CCD3D)~\cite{clay2018gpu} and Quicksilver~\cite{richards2017quicksilver} among others ~\cite{bercea2015performance, gayatri2018case,martineau2017productivity,clay2018gpu,vergara2020experiences}.  
We chose to go with OpenACC instead of the OpenMP offloading model. Based on our experience, the latter still has room for improvement before it could be more easily adopted for real-world applications, whereas OpenACC has already proven its easier adaptability, stability, and maturity. Having said that, for our continued efforts on PhysiCell in the near future, we plan to explore the OpenMP offloading model as well.

\subsection{Using OpenACC programming model for PhysiCell}
PhysiCell-GPU is designed to run the diffusion portion of PhysiCell (BioFVM \cite{ghaffarizadeh15_bioinformatics}) in an accelerated mode. BioFVM simulates the diffusive transport of substrates, while PhysiCell simulates the cells \cite{physicell}. BioFVM has been parallelized on multicore systems using directive-based programming model, OpenMP~\cite{openmp} that was established in 1998 and is widely used for parallel programming. While parallelization on several cores can be beneficial, we hypothesized that the move to the GPU would see even greater performance in key sections of the code that would benefit from being accelerated on GPUs. 

To test this, we have explored the usage of OpenACC in order to achieve performance while maintaining cross-platform portability in BioFVM. Irrespective of using OpenACC or OpenMP, the fundamental idea of a directive-based programming model for GPUs includes taking an approach similar to the low-level programming framework for GPUs, such as CUDA~\cite{cuda}, where data needed for compute is passed to the GPU and the parallel functions are computed on the GPU, and then finally pulling the updated data back to the host for further processing. 

Preliminary profiling results showed that the diffusion core was a significant portion of the total runtime; this is consistent with our earlier discussion that 
there are 600 diffusion steps ($\Delta t_\textrm{diff}$) (which requires computing updates for each voxel location in space as well as secretion and uptake for each cell agent) for every cell biology step ($\Delta t_\textrm{cell}$). Given our analysis of the parallel regions, we began to incrementally port the diffusion core of the code to GPU while carefully leaving intact biology-focused functionality in the individual cell agents, including their interactions with diffusing substrates. We started with a 2-D BioFVM test case to assess and understand the data that are required by the diffusion and supplementary functions, then began running a custom test case to cement the initial understanding. In terms of the porting process, data is handled first, then functions, then supplementary functions such as AXPY (a basic linear algebra subroutine). The overall design goal is to create a GPU accelerated diffusion portion while maintaining one-to-one equivalence of calculations between the original BioFVM and the GPU-optimized BioFVM. 

\subsection{Profiling; Identifying Hot Spots}
\begin{figure*}[h]
    \centering
    \includegraphics[width=10cm]{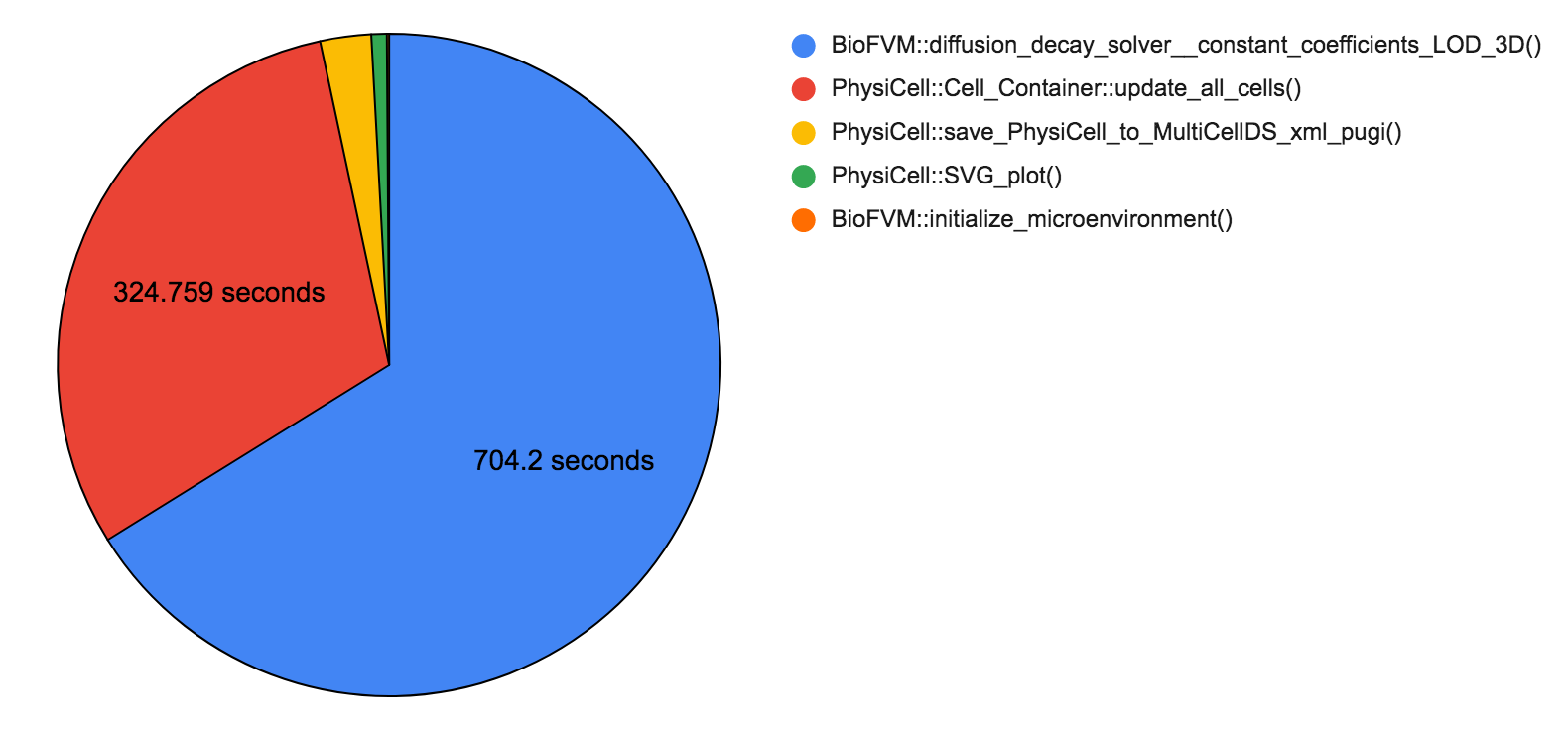}
    \caption{\textbf{Initial code profiling.} Analysis with NVProf identified that approximately 2/3 of execution time was 
    spent on biological diffusion (blue), with evaluation 
    of the discrete cell agent mechanical and biological rules (red) 
    also requiring significant computation time.}
    \label{fig:pie_chart}
\end{figure*}

We initially profiled the ``cancer-immune-sample'' sample project that is bundled with every PhysiCell download \cite{physicell}. This 3-D sample is representative of (CPU-based) PhysiCell models: the model includes multiple diffusible substrates (oxygen and an inflammatory factor), and multiple cell types (red immune cells attack tumor cells), heterogeneous cell properties (blue tumor cells are less aggressive and less immunogenic; yellow tumor cells are more aggressive but also more immunogenic), and customized cell-cell mechanical interaction rules (immune cells seek tumor cells, test for contact, adhere with spring-like terms, test tumor cell properties, and probabilistically induce death). (See Figure \ref{fig:physicell_example}.) This profiling was used in order to identify the hot spots that would be candidates for acceleration. 
% Prior to using the current NVIDIA Nsight tools, 
The NVPROF command-line profiler confirmed the insight that diffusion was the dominating portion of the total runtime. Figure \ref{fig:pie_chart} shows that the diffusion function dominated the total time, and hence was our target of optimization. For profiling, we used an internal UD system called Skywalker with an NVIDIA V100 GPU for the initial profile, and recreated the profile for the pie chart using NVPROF from Cuda release version 11.0.

%    \centering
%\includegraphics[width=8cm]{pie chart.png}
%\caption{NVProf pie chart}
%\label{fig:pie chart}
    
As we developed the OpenACC accelerated code, we periodically 
profiled the work in progress using the NVIDIA Nsight Systems profiler~\cite{nsight_systems} and an NVIDIA V100 GPU with 32 GB of High Bandwidth Memory (HBM2).

\subsection{Elements of Parallelism}
%ask Dr Macklin for good few lines about the connection between BioFVM and PhysiCell
In PhysiCell, diffusion involves a number of computational steps and function calls. The data that are invoked by these functions are handled with care to ensure a good computation vs memory transfer balance. After initialization of the GPU copy of ~\textit{p-density-vectors} in GPU memory, updates are only processed to the host if the users request (user manually managed memory) or when there is a page-fault on the data initiated by the host (OpenACC managed memory). Because the data are  ported to arrays, each part of the diffusion section is carefully mapped from its original state operating on vectors to operating on arrays. 

Data~\textit{p-density-vectors} is the main array for the data that would be updated from the diffusion functions and transferred between device and host. Originally, this data structure is a "vector of vectors" or a 2D vector (each voxel in the simulation grid holds a vector of diffusible substrate values). 
The \textit{p-density-vector} on the host are std::vectors and the p-density-vector-GPU on device are array. \textit{transfer-3D()} and \textit{transfer-2D()} are called to initialize the memory space on the GPU and to copy the data currently in \emph{p-density-vectors}. 
The functions \emph{translate-vector-to-array()} and \emph{translate-array-to-vector()} translate the data back and forth between array and vector types on the host. 
This is used after the host-side mirror of the GPU array data is updated to the host. If there is any work done in \emph{p-density-vectors} on the host, then these methods will convert the data to array form and initiate the update to the device. 

The rest of the data needed by the diffusion section are initialized in the \emph{transfer-3D()} and \emph{transfer-2D()}. These data are essential to the computation and are converted to array form like \emph{p-density-vectors}, but the data are not typically updated back and forth between host and device. If the need arises, a function could easily be made to create a host-side mirror and OpenACC directives to handle the memory updates.

The following psudeo-code shows the pattern which surrounds the OpenACC pragmas that handle data flow between the device and host. This example shows an array of arrays named physicell\_array undergo a deep copy, each inner array gets allocated on the device while keeping the pointers intact, so that the outer array can be referenced. The \emph{this} keyword is crucial to the process, as that lets the compiler know to keep the newly created data on the device interlocked with the surrounding data.

\begin{figure*}
\begin{minted}{c++}
#pragma acc enter data create
#(this->temp_physicell_array[0:bin_physicell_array][0:0])
    for (int i = 0; i < bin_physicell_array; i ++) {
        	int sze = physicell_array[i].size();
        	temp_physicell_array[i] =
        	physicell_array[i].data();
        	#pragma acc enter data copyin
        	#(this->temp_physicell_array[i:1][:sze])
    }
\end{minted}
\end{figure*}

Compute~\emph{x-diffusion-GPU-2D()} exists for the direct computation step for diffusion. In 2D, this is performed in X and Y and in 3D in X, Y, and Z vectors. The algorithm from the original PhysiCell stayed the same, while the syntax was changed to account for the new array format of the crucial data. Slight modifications are made to the original design, including creating a separate AXPY-GPU function, wrapped in an OpenACC routine directive to enable GPU execution. Storing the specific size of the \textit{p-density-vectors} in GPU memory is important for many loop bounds in the \emph{x-diffusion-GPU-2D()} section. 
\emph{apply-dirichlet-conditions-GPU()} is an important auxiliary function that was initially not set to need to run on the GPU. We learned that with the Dirichlet condition on the GPU, it eliminated the need for a high rate of update with fresh data between Dirichlet and Diffusion computation steps. The original version utilized OpenMP parallel for across the top-level loop, and our version kept to a similar organization to adhere to correctness. 

\emph{axpy-acc()} helps manage the more complex diffusion section by replacing the inline axpy and naxpy (basic linear algebra subroutines) in each x, y, and z diffusion function with a single function. Abstracting AXPY away from the multiple inline improves readability and promotes a modular design. 

\subsection{Methods of Validation for Data Integrity}
%its own full section?
We used three methods to validate that the OpenACC implementation did not affect the scientific results of the code.

\paragraph{Method 1: 1-D convergence test}
Our first validation test method uses an 1-D convergence test from the original BioFVM method paper \cite{ghaffarizadeh15_bioinformatics} to 
ensure that the expected numerical accuracy was maintained. Briefly, 
this tests the $\ell_\infty$ norm of the numerical solution against a known analytical solution at multiple times for a 1-D problem with diffusion, zero flux (Neumann) boundary conditions, and a nontrivial initial condition.  See \cite{ghaffarizadeh15_bioinformatics} for further details. 

\paragraph{Method 2: p-density-vectors cross check} 
In Method 2, we compared the data from the PhysiCell-GPU against the original CPU-based PhysiCell in each voxel. As noted previously, PhysiCell-GPU uses an array version of the data structure p-density-vector to represent the micro environment data on the GPU. PhysiCell originally used C++ std::vectors, which have an issue porting easily to GPU. In the past, there has been a limited implementation of std::vector by the PG Group, but given the complexity of PhysiCell's \textit{p-density-vector} data structure and the microenvironment C++ class structure, this implementation would not work in this case. P-density-vector is a pointer to a \verb+std::vector<std::vector<double>>+, and this extra layer prevents us from utilizing the feature. PhysiCell-GPU has a number of differences we needed to ensure did not introduce any difference against the reference. Testing for compiler, compile flag, and architecture difference was important to assess PhysiCell-GPU. 
    
\paragraph{Method 3: Visual Inspection of output} 
Methods 1 and 2 gave quantitative ways to validate PhysiCell-GPU against simple examples with known analytic solutions. We devised Method 3 to give a test against more realistic conditions in a higher dimensional geometry that more closely replicates typical modeling conditions, but where analytic solutions are not available. To verify the success of the port in terms of keeping full integrity of the data after going through diffusion on the GPU, visual inspection was used on the output for a fast yet reliable method. We had created a Python script to display the necessary data from the Microenvironment on a plot. By design, a small change in functional performance would cause a dramatic change in cell data and therefore the visualization, thus helping us to readily identify errors. 

\begin{figure*}[htb]
    \centering
    \includegraphics[width=8cm]{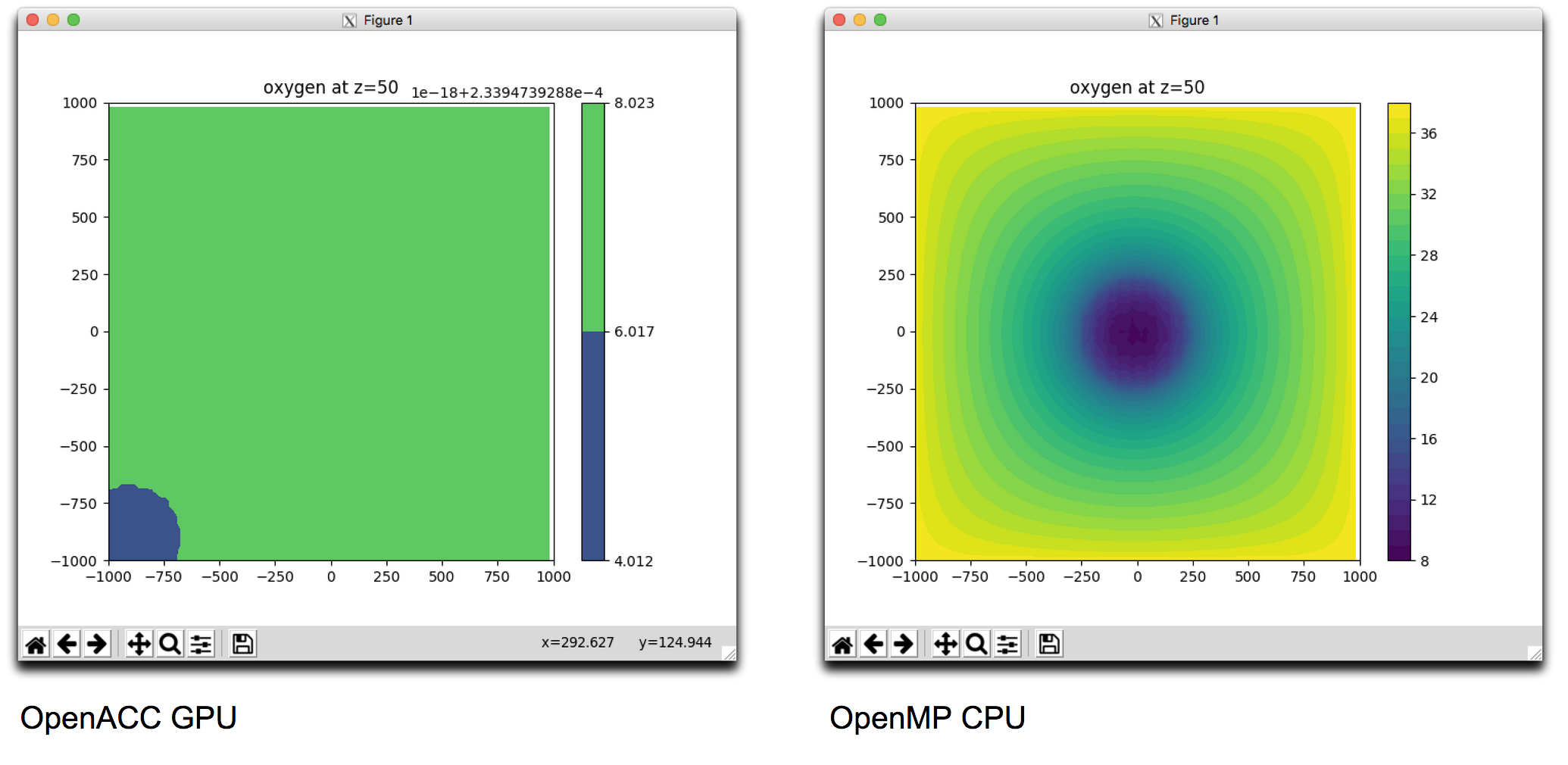}
    \caption{Method 3 validation shows a drastically different visual appearances in the presence of an ``off by one'' bug.}
    \label{fig:img1}
\end{figure*}

\begin{figure*}[htb]
    \centering
    \includegraphics[width=8cm]{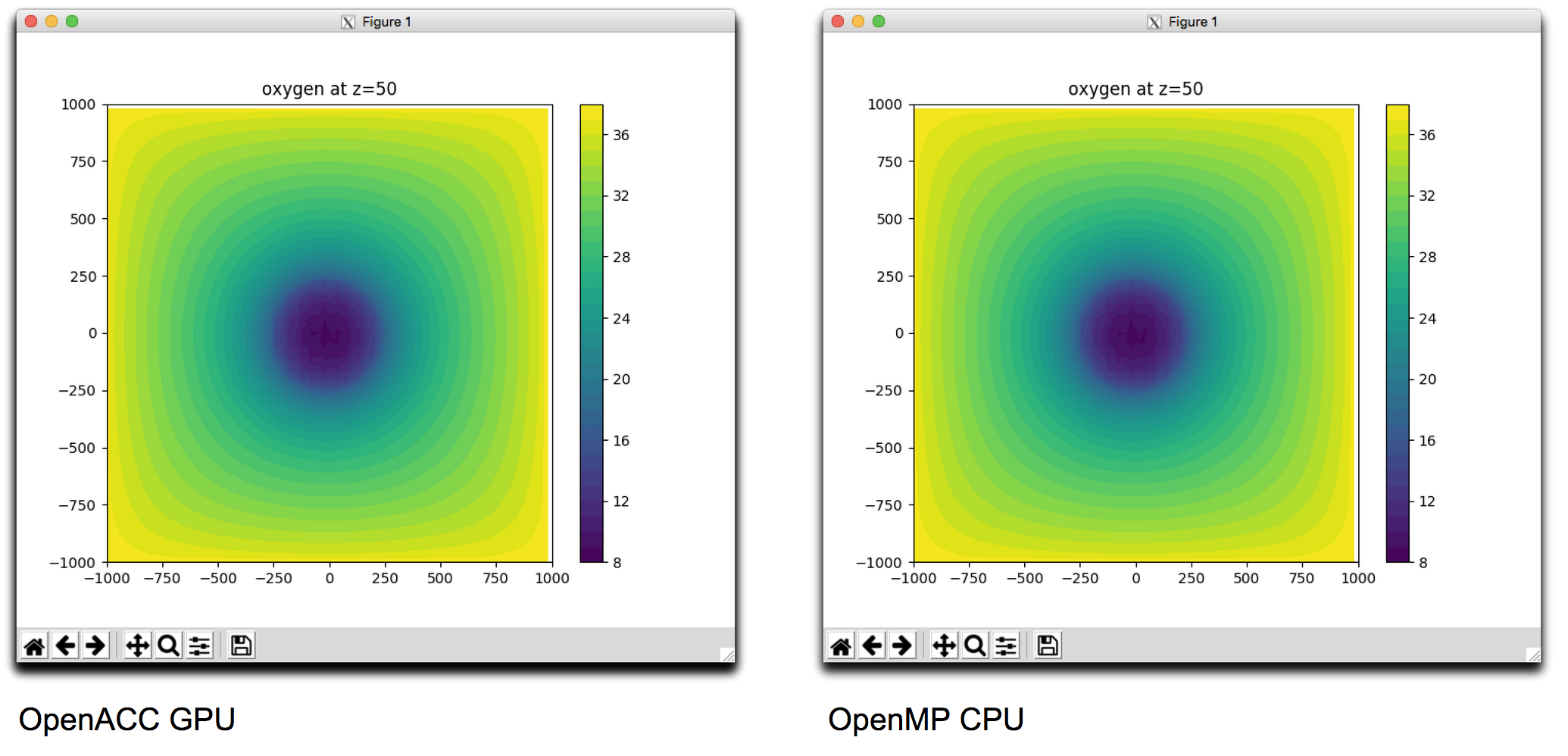}
    \caption{Method 3 validation shows identical images when implementation bugs have been fixed.}
    \label{fig:img2}
\end{figure*}

Figure \ref{fig:img1} shows an example of vastly different outputs caused by a simple ``off by one'' error in the Dirichlet function; Figure \ref{fig:img2} shows the same visual inspection test after the bug was fixed. While techniques (noted above) were used that inspected every element, the visual inspection method was often used before any other as a glaring indicator of data corruption.  
% Figure \ref{} shows two outputs that are identical, one using GPU and the other using CPU.

%best way to display these?

% Mainly Matt's todo and some bits for Robbie 
% this will be our main section to focus
% profiling and a pie plot...what routines/functions stood out 
% explain the routines a bit - what is their functionality
% a paragraph on OpenACC execution and memory model - Robbie could take a stab
% Porting process - how did you work with functions/loops/data structures and how did you add openacc to it, what were the directives you used and how you used them. OK to add 1-2 pseudo snippets. Definitely refer to the PLOS one journal (PPM paper by Eric) for the style 
% did we re-profile and improve the function further? did that improve the overall codebase 
% limitations of the port? 

\section{Experimental Setup and Input Datasets}
This section highlights the experimental set up details for the results. We have primarily used two NVIDIA DGX machines. The hardware and the software details are described below in Table~\ref{tab:machine-specs}. 

\begin{table*}[h]
  \centering
  \scalebox{0.9}{
  \begin{tabular}{|c|c|c|}
    \hline
    Machine & CPU & NVIDIA GPU \\ \hline
    NVIDIA DGX-2 & Intel Xeon Platinum 8168 (24 cores) & Volta V100 (32GB HBM2) \\ \hline
    NVIDIA DGX A100 & AMD EPYC Rome 7742 (64 cores) & Ampere A100 (40GB HBM2) \\ \hline
    
  \end{tabular}}
  \caption{Specifications of the nodes in the two systems}
%     \vspace{-2em}
  \label{tab:machine-specs}
\end{table*}

While the original PhysiCell code was compiled using OpenMP-enabled $gcc$ version 7.5.0, we used the NVIDIA HPC SDK's $nvc$ 21.3 compiler for parallelizing and accelerating the code. This compiler supports both OpenMP parallelization, as well as OpenACC offloading. The full code corresponding to this paper is available on GitHub~\cite{physigithub}. 

%To compile and parallelize the code on the CPUs, we have used NVIDIA HPC SDK OpenMP, NVHPC 21.3 compiler. 
%To compile and accelerate the code on the GPUs, we have used 
%NVIDIA HPC SDK OpenACC compiler version 21.3. 
%The OpenACC compiler and version used are the same for both CPUs and GPUs. 
%We have also compiled and parallelized the code on the CPUs, using the same OpenACC compilers. 

\subsection{Input Datasets}
PhysiCell input is written in XML. In the original PhysiCell code the parameters changed were x min, x max, y min, y max, z min, z max, max time, and omp num threads. The set of results came from x,y,and z min set at -1000 along with x,y, and z max set at 1000 with the max time units at 60 simulated minutes. The parameter omp num threads was set at 1 and 32, representing a serial implementation and a optimal performance CPU thread count respectively. The second set of results kept all parameters the same expect increasing the max time units to 180 simulated time, and a third for 360 simulated minutes. This input set represent a cell at the size of 1cm. 

%\red{set at 1 and 32 only, or varied from 1 to 32?} \blue{Since this run is on a single socket, we are good with showing results for 32 cores, if MPI we can show a nice scaling}
\section{Results}
This section presents the results on the CPUs and the GPUs. Table~\ref{tab:results} presents results for single-core CPU, 64 cores CPU, NVIDIA's V100 GPU, and NVIDIA's A100 GPU using two implementations namely the manual and the managed memory implementations for the different input datasets.

\begin{table*}[h]
  \centering
  \scalebox{0.9}{
  \begin{tabular}{|c|c|c|c|c|c|c|}
    \hline
    \textbf{Sim Dataset} & \textit{60 Sim mins} & \textit{180 Sim mins} & \textit{360 Sim mins} \\ \hline
     \textbf{OMP CPU 1 Core} & 8 min. 44.6083s & 25 min. 11.1268s & 51 min. 47.043s \\ \hline
     \textbf{OMP CPU 64 Cores} & 1 min. 6.0669s & 3 min. 21.9457s & 6 min. 44.9028s \\ \hline
     \textbf{ACC CPU 64 Cores} & 57.993s & 2 min. 47.4116s & 5 min. 30.3994s \\ \hline
     \textbf{Manual GPU V100} & 1 min. 34.2378s & 2 min. 39.4965s & 4 min. 17.9657s \\ \hline
     \textbf{Manual GPU A100} & 2 min. 20.6413s & 3 min. 36.9927s & 5 min. 25.707s \\ \hline
     \textbf{Managed GPU V100} & 23.903s & 57.4191s & 1 min. 47.7914s \\ \hline
     \textbf{Managed GPU A100} & 21.3251s & 45.9034s & 1 min. 22.7607s \\ \hline
  \end{tabular}}
  \caption{Results Table}
%     \vspace{-2em}
  \label{tab:results}
\end{table*}

%\subsection{Experimental setup}

%Further details for the UDEL's Skywalker includes the AMD Ryzen Threadripper being a 16-core cpu, 47GB of DDR4 ram, and a Nvidia Tesla V100 16GB GPU with compute capabilities 7.0; running Ubuntu 18.04.4 LTS. The OpenACC version of PhysiCell was compiled with PGI compiler version 19.10. While the original PhysiCell code with OpenMP was compiled with g++ version 7.5.0.\\

%The University of Delaware's Skywalker machine was used for the results of the OpenMP CPU tests and the OpenACC GPU tests with V100. The CPU in this machine is a AMD ThreadRipper 1950X (16 cores/32 threads) with 48GB RAM. The GPU is a NVIDIA Tesla V100 with 16GB HBM2. 

\begin{figure*}[h]
    \centering
    \includegraphics[width=13cm]{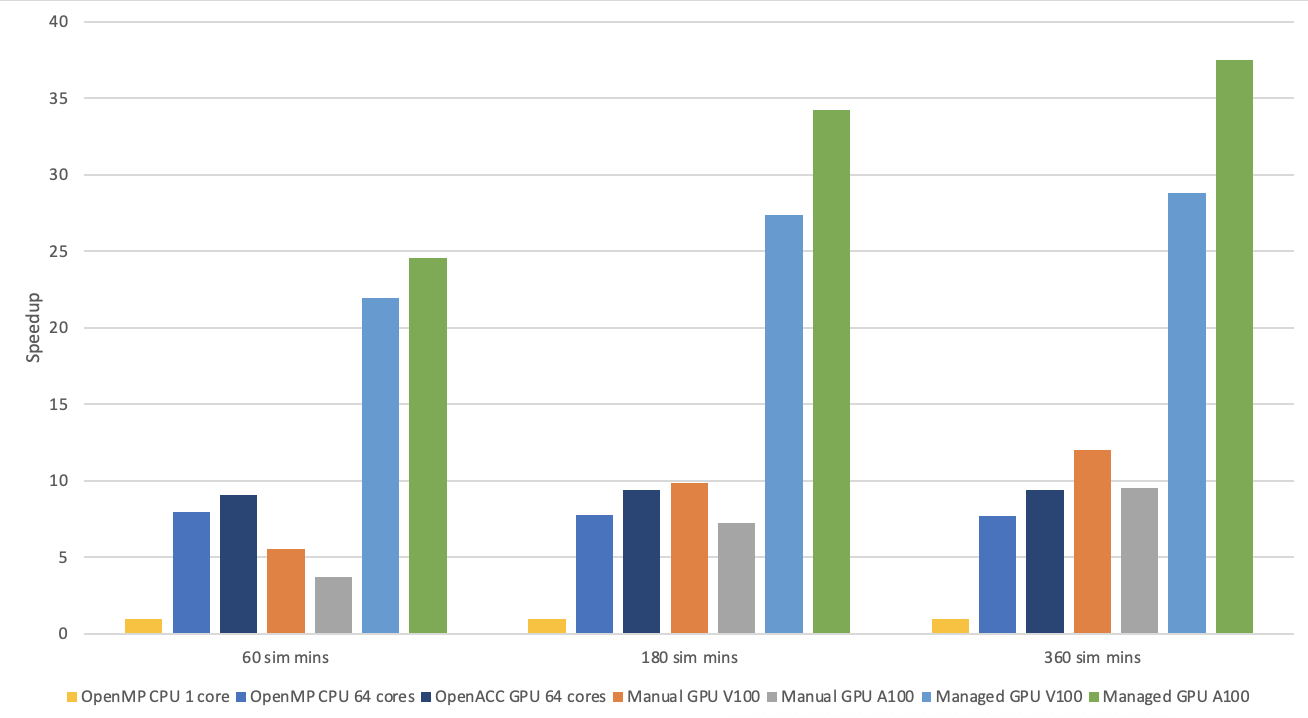}
    \caption{Speedup normalized over serial }
    \label{fig:Results Table}
\end{figure*}

The OpenMP CPU results use NVIDIA's NVHPC 21.3 showing the time taken using 1 thread on AMD EPYC Rome to simulate a serial version. Using 1 thread is close to running the code in the native serial manner, but it is not exact, as the code has been optimized from the native serial version. 
We were forced to take this approach because a native serial version of the code is not available. 
We then show the runtime using 64 cores to demonstrate the speedup of approximately 7.6x when running the OpenMP version on a high end CPU.
Figure~\ref{fig:Results Table} shows that a speedup of almost 40x using OpenACC to target an NVIDIA A100 GPU using managed memory was achieved over the serial CPU configuration for 360 simulated minutes as the dataset. 
Overall, the NVIDIA A100 GPU shows a better speedup with the same code over the V100 GPU because of the more advanced architecture and faster memory interconnect of the A100s. 

\begin{comment}
A100 Machine (NVIDIA's DGX A100): \\
cpu-AMD EPYC 7742 64-Core Processor\\
gpu-A100-SXM4-40GB\\
V100 Machine (UDEL CRPL's Skywalker):\\
cpu-AMD ThreadRipper 1950X (16 cores/32 threads), 48GB RAM\\
gpu-Tesla V100 (16GB HBM2)\\
\end{comment}

% normalization by the single core --- to show the orders of magnitude... 

%\subsection{Performance Impact}

The table also shows the execution time of the manual and the managed memory implementations for the GPUs. The manual implementation copies the data back and forth between the host and device using explicit OpenACC data management directives and translation between vector and array data types.
The managed memory results refer to compiling the code with the managed memory flag turned on, which is similar to CUDA's managed memory implementation; programmer intervention with respect to data movement is not required. 
The addresses in host and device point to the same address location, and use page-faults to trigger data updates/migrations.
%The manual memory implementation has a copy of data on the GPU, and a separate copy on the CPU, requiring explicit data copy and transfer between the GPU device and CPU host. 
%The managed memory implementation has one copy of the data, and it is able to update to the GPU or the CPU with the latest data, managed by the compiler without programmers input. 
In our case, the managed memory implementation has the benefit of less memory transfer time compared to manual memory implementation in exchange for less fine-grain control of the GPU resident data. This relationship can be observed in Figure~\ref{fig:Results Table} in the form of relative speedups. 

\begin{figure*}[htb]
    \centering
    \includegraphics[width=12cm]{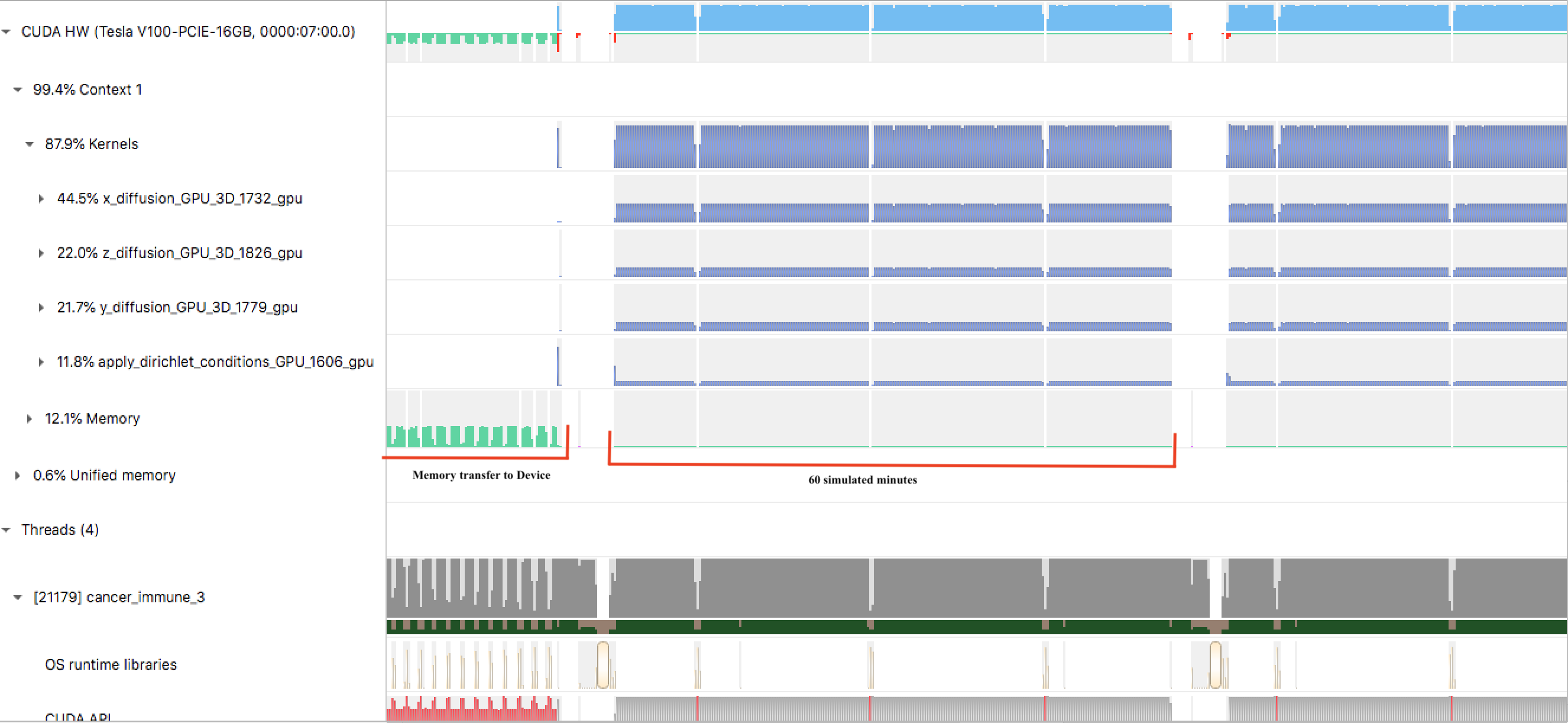}
    \caption{NVIDIA Nsight Systems Profile of Physicell\_GPU with managed memory}
    \label{fig:profile_managed}
\end{figure*}

This shows that while data transfer from the CPU to the GPU is relatively expensive upfront, the benefit is seen over more timesteps with more realistic simulation lengths. This is due to the fact that data can remain on the GPU in between timesteps, which reduces the amortized cost of the upfront data transfers and increases the benefit of the accelerated diffusion as the number of timesteps is increased. Figure~\ref{fig:profile_managed} shows an Nsight Systems profile of the code running on the GPU using managed memory. We observe the large upfront data transfer cost in green, followed by the computation that makes up each timestep in blue. Note that there are no large pieces of data movement to or from the GPU in between the timesteps shown.
Overall, we see speedups that benefit the domain application by allowing larger problem sizes and longer simulation times in less wall time. 

The times for the Manual GPU A100 run show that there was a greater memory transfer time than the Manual GPU V100 run. This was unexpected and not caused by the A100 GPU itself, as we have tested on a standalone A100 PCIE based card and saw no extra memory time. We believe the extra minute was caused by user issue with the Slurm job scheduler, due to the fact that the CUDA memory bandwidth test showed expected high speed. Regardless of the unexpected extra transfer time, the speed of the diffusion was faster than the Manual GPU V100 as expected for the faster chip, showing only an increased time in transfer. The DGX A100 was used for consistency across test version.

%Figure~\ref{fig:Results Table} shows the speedup across the board normalized over the serial execution. 

This is the first work to apply portable GPU acceleration (via OPenACC) to biological diffusion in PhysiCell--a critical bottleneck to larger and longer simulations. Others have recently applied MPI accelerations to BioFVM \cite{biofvm_X} and PhysiCell to advance towards billion-cell simulations \cite{gigascale_ABM}. However, these
require significant code refactoring and high performance computing resources 
to attain their full performance. They have not been optimized to leverage workstation (or single compute node) GPU resources. Ultimately, a hybrid OpenACC-MPI architecture 
could attain still better performance by leveraging our OpenACC-based GPU acceleration on individual compute nodes, with MPI and domain decomposition to attain billion-cell 
scalability.
% \red{``first'' statements prob belong in supplementary materials. 
% Barcelona folks are doing MPI acceleration but not GPU or OpenACC.''} 
% \blue{SC: Paul, okay, wanted to say something along the lines of this is the first of % that equation 1 in PhysiCell on GPU... I didn't exactly want to say 'first' work but % not sure how to word it}

\subsection{GPU-accelerated Long-time Simulations}
To assess the potential scientific impact of the GPU-accelerated code, we 
performed tests on PhysiCell's built-in \verb+cancer_immune_sample+ sample project that originated in 
\cite{physicell}. (See the brief description of this sample project 
in Figure \ref{fig:physicell_example}.) To maintain our focus on the GPU 
acceleration of the diffusion code, we disabled cell mechanics and phenotype 
changes, thus leaving cell positions, sizes, and behaviors fixed. To emulate a typical scientific workflow, we performed a full data save every 7 simulated days. In the original work \cite{physicell} and subsequent 
3D parameter space investigations \cite{EMEWS1}, simulations were 
limited to 21 days (3 data checkpoints) dues to CPU limitations, each requiring approximately 48 hours (wall time) to complete. Based on our profiling, we 
estimate that diffusion consumed approximately 65\% ($\sim$30 hours) of the wall time for each simulation in prior studies.  
%Very rough estimate: New code is 40x faster on diffusio, which is 50\% of the compute time. %So, new compute time is $\frac{1}{2} + \frac{1}{40} \frac{1}{2} = \frac{41}{80}$. 

After recompiling the CPU-based code with the same \texttt{nvc++} compiler, 
we simulated the 3-D model (diffusion only) with default parameter values 
on an AMD EPYC 64-core CPU.  
% To emulate a typical scientific workflow, we performed a full data save every 7 simulated days. 
The CPU-only code only reached the first checkpoint (7 days) and 
was not able to complete the full 21-day simulation within the 
maximum 4-hour time limit on the DGX system (a typical constraint on large-scale 
model exploration on shared systems); based on the simulation's progress by this 
time limit and prior benchmarking \cite{ghaffarizadeh15_bioinformatics}, 
we estimate that the CPU-only code would require approximately 
9.3 hours (wall time) to complete the full 21-day simulation on the system.  
%\red{(Sunita and Matt: If the 64-core CPU simulation timed out at 4 hours after simulating 7 days, how do you estimate that 8 hours and not 12 are required for the full 21-day 
%simulation? I tweaked the estimate to 9.3 hours based on your figures below.)}
Only the first checkpoint (equal to a 7 day simulation) finished at 3 hours, 6 minutes and 4 seconds of wall time. 
%\red{(11164 seconds per week, 
%so 33492 seconds for 21 days, or 9:18:12.)} 
In comparison, the OpenACC GPU-accelerated code (managed memory, A100) completed the entire 21-day simulation (3 checkpoints) in 1 hour, 42 minutes and 13.52 seconds. 

%The extended simulation---made feasible with the GPU accelerations, showed previously 
%undetected dynamics .. \red{describe new dynamics}. See Figure \ref{fig:extended_simulation}.

We next asked how far an OpenACC GPU-accelerated simulation (managed memory, A100 GPU) 
could be extended within a maximum 4-hour time limit by increasing the test's maximum simulation time and continuing to save data every 7 simulated days. Whereas the CPU-only code (64 cores; AMD EPYC) only reached the first 7-day checkpoint, the OpenACC code was able to reach and pass the 42-day checkpoint (6 checkpoints), requiring 3 hours, 24 minutes to simulate 42 days of wall time. By contrast, 
we estimate that CPU-only code would have required over 18 hours to complete the 
same computations. 
% \red{continuing here}
% Using $nvc++$ to target the CPU using OpenMP and 64 cores to simulate 7 days, i.e. just 1 time step interval, the code took 3 hours, 6 minutes and 4 seconds. 
% CPU: 11164 seconds / checkpoint
% CPU: 66984 seconds for 6 checkpoints = 18 h 36 min 24 sec 
This shows that using the GPU-accelerated diffusion code should enable long-duration simulations (e.g., of cancer and disease progression) 
that were previously only feasible on systems that permit submitted jobs to run for hours or days. However, many shared computational resources are not available for the length of time that would be required. Without accelerated computing, a timeout of resources would be reached long before the desired results were computed.

This is particularly relevant for the emerging field of 
cancer forecasting \cite{forecasting_cancer} and 
cancer patient digital twins \cite{precision_oncology,nature_medicine_CPDT}: 
after calibration, a clinical teams will use the patient's digital 
twin to simulate thousands of candidate treatment plans (virtual control, 
standard of care, and multiple combination therapies under a variety of 
dosing options) over weeks or months, each with tens of replicates to estimate uncertainty. 
These multi-week simulations must execute quickly to allow timely clinical 
decision support; supporting clinical workflows \emph{at scale} will 
further increase the need for rapidly executing long-duration forecasting 
simulations. 

% \begin{figure}[htb]
% \includegraphics[width=\0.99\textwidth]{extended_results.png}
% \caption{\textbf{Extended cancer-immune simulation.} The GPU-accelerated 
% code allowed us to extend 3-D cancer-immune simulation to 42 days with 
% approximately the same computational time as a 21 day simulation 
% in the original CPU code. See the extended simulation dynamics 
% at 21, 28, 35, and 42 days.}
% \label{fig:extended_simulation}
% \end{figure}

\begin{comment}

\vspace{2in}  
\red{[experiments to do 
Let's do one last example, if we can: run the cancer-immune-sample with default parameters, except:]} \blue{SC: I like this, worth a try, will ping Matt}\\

Run A (CPU version)
\begin{enumerate}
\item Set \verb+max_time = 30240+ (21 days = 3 weeks)
\item Set \verb+save+ $\rightarrow$ \verb+full_data+ $\rightarrow$ \verb+interval+ to 10080 (7 days)
   \item Set \verb+save_interval_after_therapy_start = 1440+ 
\end{enumerate}
How long does A take? 

Run B (best GPU version)
\begin{enumerate}
\item Set \verb+max_time = 60480+ (42 days = 6 weeks)
\item Set \verb+save+ $\rightarrow$ \verb+full_data+ $\rightarrow$ \verb+interval+ to 10080 (7 days)     \item Set \verb+save_interval_after_therapy_start = 1440+ 
\end{enumerate}
How long does B take to reach 21 days? Is it ~1/2 of the simulation time for A? 

How long does B take to reach 42 days? Is it less than 21 days in the CPU version or 
about the same? 

\red{end description} 
\end{comment}

% add discussion of figure 5
\section{Conclusion}

Code profiling of a complex, multiscale biological agent-based code revealed that for typical simulation models, approximately 65\% of execution time (wall time) is spent on biological diffusion, making this a logical first target for GPU optimization. Offloading numerical operations to solve the diffusion along with careful management of memory transfers between host and device memory resulted in an approximately 40-fold reduction in execution time for the biological diffusion solver. Multiple testing methods were used to validate the GPU code against the original CPU code. 

In real-world testing against a complex 3-D cancer immunology example, switching from a 64-core CPU implementation to a managed memory OpenACC GPU implementation reduced execution time (wall time) for 21 days of biological diffusion 
and data saves from 9.3 hours to 1.7 hours---a reduction 
of over 80\%. Based on earlier profiling that 
diffusion is responsible for over 65\% of total 
execution time, we estimate that full simulation 
code---once again include cell mechanics and biological 
calculations responsible for the remaining 35\% of 
execution time---would require approximately 50\% 
less execution time once using the GPU-accelerated 
diffusion algorithms developed in this paper. 
Furthermore, the GPU-accelerated code was able to complete a diffusion-only simulation twice as long as the CPU-based code in roughly one third of the time, allowing us to simulate farther in time than was previously feasible while observing new biological dynamics. 

% total_old = diff + non_diff
% total_old = 0.65 * tot_old + 0.35 * tot_old 
% 
% tot_new = 0.65*0.2 * tot_old + 0.35 * tot_old 
% tot_new = (0.65*.2 + 0.35 ) * tot_old 
% tot_new ~ 0.47 * tot_old ~ 0.5 * tot_old 

% This confirms the scientific benefit of accelerating targeted portions of the code with careful GPU optimization. %% CPU : 21 days in 9.3 hours
%% GPU: 21 days in 1.7 hours
%% GPU : 42 days in 3.4 hours 
% 3.4 / 9.3 = 0.365 

A full 21-day simulation might be expected 
to take $\sim\hspace{-1.6mm}\frac{1}{0.65} \times$ 9.3 
hours (14-15 hours) to execute, 
and so the GPU-accelerated full model 
should see its total wall time 
reduced to 7-8 hours. This could potentially 
allow us to double the duration of each simulation 
for the same amount of execution time, thus 
exposing new biological dynamics. 
This confirms the scientific benefit of accelerating targeted portions of the code with careful GPU optimization. 
%
% t_old = GPU + nonGPU 
% GPU = f * t_old = f * (GPU + nonGPU)
% (1-f)*GPU = f*nonGPU 
% nonGPU = (1-f)/f * GPU 
% f = 0.65
% t_new = 0.2 * GPU + nonGpU 
% t_new = 0.2 * GPU + (1-f)/f * GPU 
% t_new = (0.2 + (1-f)/f ) * GPU
% 
% t_new / t_old = (0.2 + (1-f)/f ) / ( 1 + (1-f)/f) 
% f = 0.65
% (0.2 + 0.54 ) / ( 1 + .54 ) = .74 / 1.54
% 9.3 hours --> 1.7 hours for 3 weeks
% 186 min --> 34 minutes for 1 week. 
%
% nonGPU for 7 days: 
% t_old = GPU + nonGPU 
%       = GPU + (1-f)/f * GPU 
%       = (1 + (1-f)/f) * GPU 
%       = (1/f) * GPU 
% (.65)/.35 * 9.3 hours = 17 hours 

\section{Acknowledgement}
This material is based upon work supported by the National
Science Foundation under Grant No. 1814609.

\section{Future Work}

The current implementation yielded a significant performance increase that in turn ``unlocked'' new scientific possibilities in the code, particularly longer-time simulations. However, more could be done. 
First, the OpenACC acceleration of the biological diffusion still relies upon CPU operations for the cell-based secretion and uptake of biological substrates. These may account for the difference between the 40$\times$ speedup in simpler performance benchmarking and the ``real world'' 3D test (5-6$\times$ speedup) where cell secretion and uptake play a greater role. Moving these operation on the GPU could further reduce the need for costly memory transfers, allowing all computations to ``reside'' on the device for 10 computational diffusion steps (with size $\Delta t_\textrm{diff} \sim 0.01\textrm{ min}$) without need for memory transfer. Second, cell mechanics operations are governed by biased random migration and interaction potentials, which are well-suited to GPU computations~\cite{wright2020accelerating}. Furthermore, there are 60 mechanics steps (with step size 
$\Delta t\textrm{mech} \sim 0.1 \textrm{ min}$) for every cell step (with size 
$\Delta t_\textrm{cell} \sim 6 \textrm{ min}$). Thus, moving cell mechanics solvers  
would not only accelerate those computations, but drastically reduce the the need for 
host-device memory transfers: computations could ``reside'' on the device for 600 computational 
steps before transferring data to host memory. 

%
% t_old = diff + mechanics + rest
%          65%  + 25%  + 10% 
% t_new ~ (1/40)* 0.9 + 0.1 
% t_new ~ (0.0225  + .1 )  * 15.3 ~ 2 h

Future work will explore these refinements. 
Biological diffusion---which present work accelerated 
by a factor of 40---accounts for approximately 
65\% of execution time in typical simulations. 
Cell updates---dominated by 10 mechanics steps 
for every biological step---accounts for another 25-30\% 
of execution time. If the 
mechanics code (approximately 25\% of original execution time) could be similarly 
accelerated by a factor of 40, 
% (i.e., 
% if 90\% of the original code execution 
% time could be reduced by a factor of 40), 
then the overall simulation execution time should 
be reduced on the order of 85-90\%: a substantial 
speedup. 

A 10$\times$ to 100$\times$ speedup would enable exciting 
new scientific possibilities. If simulation sizes were left unchanged, then the speedup would enable much faster execution times for individual simulations; this is critical for data assimilation and parameter estimation techniques like approximate Bayesian computing that must rapidly run many simulations sequentially. Similarly, leaving the simulation size unchanged, we could simulate to longer times, which will be of great use to digital twin efforts in medicine. If execution time were left unchanged, then larger domains with more agents could be simulated, enabling studies of more complex tissues and even small organisms. Lastly, the increased ``computational budget'' afforded by a more complete GPU acceleration could allow us to introduce multiple agents per biological cell allowing 
sophisticated simulations of not only cell morphology (e.g., as in subcellular element models 
\cite{JCO-review,sem_pp}), but even introducing agents for subcellular components and biophysical processes, 
such as movement, fission, and fusion of Golgi bodies 
during signaling \cite{Scotte2021571118,padmini2}. 
This could be transformative in relating emerging high-resolution microsopy 
\cite{whole_cell,allen}  to intracellular biophysics and functional biology. 

\bibliography{main}
\bibliographystyle{unsrt}
\end{document}